\documentclass[aps,prl,twocolumn,superscriptaddress,preprintnumbers]{revtex4-1}
\usepackage{amsfonts,amsmath,amssymb,bm,bbm}
\usepackage{bm}
\usepackage[colorlinks=true,linkcolor=blue,citecolor=blue,urlcolor=blue]{hyperref}
\usepackage{dcolumn}
\usepackage{epsfig}
\usepackage{enumerate}
\usepackage{float}
\usepackage{graphicx}
\usepackage{hhline,multirow}
\usepackage{latexsym}
\usepackage{mathrsfs}
\usepackage{mathtools}
\usepackage{physics}
\usepackage{slashed}
\usepackage{soul}
\usepackage{subfigure}
\usepackage{url}
\usepackage{xcolor}

\begin{document}
\preprint{JLAB-THY-25-4569}

\title{First simultaneous global QCD analysis of kaon and pion \\ parton~distributions with lattice QCD constraints}

\author{P. C. Barry}
\affiliation{Physics Division, Argonne National Laboratory, Lemont,             Illinois 60439, USA}
\author{Chueng-Ryong Ji}
\affiliation{\mbox{Department of Physics and Astronomy, North Carolina State University, Raleigh, North Carolina 27695, USA}}
\author{W. Melnitchouk}
\author{N. Sato}
\affiliation{Jefferson Lab, Newport News, Virginia 23606, USA}
\author{Fernanda Steffens}
\affiliation{Helmholtz-Institut für Strahlen- und Kernphysik, University of Bonn, Germany}
\affiliation{Bethe Center for Theoretical Physics, University of Bonn, Germany  \\
        \vspace*{0.2cm}
        {\bf JAM Collaboration \\ {\footnotesize (Pion PDF Analysis Group)}}
        \vspace*{0.2cm} }

\begin{abstract}
We perform the first simultaneous global QCD analysis of pion and kaon parton distribution functions (PDFs), constrained by pion- and kaon-induced Drell-Yan (DY) and leading neutron electroproduction data, together with lattice QCD data on pion and kaon PDF moments. The analysis indicates a softer valence $\bar u$ distribution in the $K^-$ than in the $\pi^-$, and a significantly more peaked valence $s$-quark density in $K^-$ compared with the $\bar u$. The effective exponent governing the high-$x$ behavior of the PDF is found to be larger for $\bar u$ in the kaon, $\beta_{\bar u}^{K^-}\!= 1.6(2)$, than in the pion, $\beta_{\bar u}^{\pi^-}\!= 1.16(4)$, in the range $0.7 \leq x \leq 0.95$. From the gluon momentum fractions we find the pion's gluon content accounts for $\approx 1/3$ of the mass budget of the pion at $\mu=2~{\rm GeV}$, but only $\approx 1/4$ for the kaon.
\end{abstract}

\date{\today}
\maketitle

{\it Introduction---}
As the lightest composite particles observed in nature, the pseudoscalar mesons---pions and kaons---play a special role in understanding how strong nuclear interactions are related to the fundamental properties of QCD, such as SU(2) and SU(3) chiral symmetry breaking, and hadron mass generation.
A central challenge has been to synthesize this role with the fact that in QCD these are also bound states of quarks and gluons, whose partonic substructure is revealed in high-energy scattering experiments such as deep-inelastic scattering (DIS) and Drell-Yan (DY) lepton-pair production. 
At the same time, the short lifetimes and absence of fixed meson targets has made it very difficult to determine their partonic structure directly.

Historically, data from pion-induced DY experiments 
\cite{Branson:1977ci, Anderson:1979tt, Barate:1979da, Corden:1980xf, McEwen:1982fe, Greenlee:1985gd, NA10:1985ibr, Conway:1989fs} have provided constraints on the valence quark distributions at parton momentum fractions $x \gtrsim 0.2$.
To access the small-$x$ region, leading neutron (LN) electroproduction from HERA~\cite{H1:2010hym, ZEUS:2002gig} was studied in terms of the Sullivan process~\cite{Sullivan:1971kd}, in which the proton splits into a $\pi^+$ and neutron, and the detection of a forward neutron allows an interpretation in terms of DIS from a nearly on-shell pion~\cite{Thomas:1983fh, McKenney:2015xis, Salamu:2014pka}.
Using the Bayesian Monte Carlo methodology developed for the analysis of proton PDFs, the JAM Collaboration has performed a number of studies of pion parton distribution functions (PDFs), incorporating LN data to more reliably constrain small-$x$ pion PDFs~\cite{Barry:2018ort}, as well as investigating impacts of the transverse momentum dependent DY data~\cite{Cao:2021aci, Barry:2023qqh}, including threshold resummation in the DY process~\cite{Barry:2021osv}, and exploring constraints from lattice QCD data~\cite{Barry:2022aix, Good:2025nny}.

For the kaon, a much smaller set of data is available, essentially from the NA3 experiment at CERN~\cite{NA3:1980fhh}, which collected 700 dilepton events from $K^-$ scattering off a platinum (Pt) target, and presented these as a ratio of kaon- to pion-induced DY differential cross sections.
To isolate the valence $\bar{u}$ PDF in the $K^-$, the kinematics of the experiment were restricted to the high-$x$ region of the ``nucleon'' (Pt) target, where the cross section is expected to be dominated by valence quarks, and contributions from the $s^K \bar s^A$ channel are suppressed.
In the same kinematics, the $\pi^-$-induced cross section should be dominated by the $\bar{u}^\pi u^A$ channel, so the measured ratio was interpreted as a ratio of $\bar{u}$ PDFs in $K^-$ to $\pi^-$.

While seemingly reasonable, the expectation of the $\bar{u}$ dominance of the $K^-$ cross section requires assumptions about the behavior of the sea quark distributions of the beam and target, so that the ratio data alone cannot uniquely determine even the flavor structure of the valence kaon PDFs.
Since the $s$ quark is much heavier than the $\bar u$, we may expect differences between the valence structures of the kaon, which unfortunately have never been identified.

In the absence of empirical information, to separate the quark flavor PDFs in the kaon, 
lattice QCD simulations can be used to complement the experimental measurements.
Calculations have been made for several low PDF moments~\cite{Martinelli:1987zd, Best:1997qp, Detmold:2003rq, Alexandrou:2021mmi, Guagnelli:2004ga, Capitani:2005jp, Abdel-Rehim:2015owa, Oehm:2018jvm, Alexandrou:2020gxs, ExtendedTwistedMass:2021rdx, Loffler:2021afv, Alexandrou:2021mmi, Hackett:2023nkr, Good:2023ecp, ExtendedTwistedMass:2024kjf, Francis:2025rya}, as well as quark quasi-PDFs~\cite{Zhang:2018nsy, Izubuchi:2019lyk, Lin:2020ssv, Gao:2020ito, Gao:2022iex,Sufian:2019bol, Gao:2021dbh}, pseudo-PDFs~\cite{Joo:2019bzr}, and current-current correlators~\cite{Sufian:2020vzb}.
Specifically, for the $K^-$, valence $\bar{u}$ and $s$ quark~\cite{Lin:2020ssv} and gluon quasi-PDFs have been computed~\cite{Salas-Chavira:2021wui}.
Recently state-of-the-art momentum fraction calculations have been performed by the ETM Collaboration (ETMC)~\cite{ExtendedTwistedMass:2024kjf} for the pion and kaon  at physical quark masses and with a continuum extrapolation, reducing the lattice systematic effects.
Higher moments, computed at slightly larger quark masses~\cite{Alexandrou:2021mmi},  can also help constrain the $x$ dependence of the PDFs.

Earlier extractions of kaon PDFs from the NA3 data were performed using constituent quark model-inspired relations~\cite{Gluck:1997ww}, as well as a statistical model for the PDFs~\cite{Bourrely:2023yzi} and employing additional constraints from $J/\psi$ production.
Recently, combined experiment and lattice analyses have become more common in extractions of nonperturbative functions, such as the Collins-Soper kernel~\cite{Avkhadiev:2025wps}, generalized parton distributions~\cite{Guo:2025muf}, helicity PDFs \cite{Bringewatt:2020ixn, Karpie:2023nyg, Hunt-Smith:2024khs, Cocuzza:2025qvf}, and pion PDFs~\cite{Barry:2022aix, Good:2025nny}.
In this Letter, we present the first Monte Carlo QCD analysis of the combined experimental cross sections and lattice moments to determine the valence quark PDFs in the pion and kaon, along with their impact on the gluon momentum fractions.
Unlike previous studies, we use a flexible, model-independent parametrization, together with rigorous uncertainty quantification.
Furthermore, since the NA3 experiment~\cite{NA3:1980fhh} measured a ratio of kaon to pion cross sections, to avoid injecting bias into the analysis from using fixed pion PDFs, we extract both the kaon and pion PDFs simultaneously.

{\it Experimental and lattice QCD constraints---}
The DY dimuon production process, $h_1 h_2 \to \mu^+ \mu^- X$, has traditionally been the main source of information for meson PDFs.
The scattering of a meson $h_1$ ($\pi^-$ or $K^-$) from a nuclear target $h_2$ is given by the factorized cross section
\begin{equation}
    \frac{\dd\sigma}{\dd x_1 \dd x_2} = \frac{4\pi \alpha^2}{9 Q^2} \sum_{i,j} 
    \big[ C_{ij} \otimes f_{i/h_1} \otimes f_{j/h_2} \big](x_1,x_2;\mu^2),
    \label{eq.DYxsec}
\end{equation}
where $C_{ij}$ are the hard coefficients describing the partonic subprocess, computed at next-to-leading order (NLO) and next-to-leading logarithmic (NLL) accuracy using the double Mellin method~\cite{Westmark:2017uig, Barry:2021osv}, $Q^2$ is the mass squared of the virtual photon, and the sum is over all partons $i,j$.
The PDF $f_{i/h_1}$ ($f_{j/h_2}$) for parton flavor $i$~($j$) in the beam (target) hadron is a function of parton momentum fraction $x_{1(2)} = (Q/\sqrt{s})\, e^{+(-)y}$, where $s$ is the total invariant mass squared of the reaction, $y$ is the rapidity, and $\mu$ is the factorization scale.

For the kaon, the NA3 experiment~\cite{NA3:1980fhh} measured the ratio $\dd\sigma/\dd x_1$ of $K^-$ to $\pi^-$ DY cross sections on a Pt target (6 data points), for $4.1 \leq Q \leq 8.5~{\rm GeV}$, with $x_2$ integrated over the range $Q_{\rm min}^2/s x_1 \leq x_2 \leq Q_{\rm max}^2/s x_1$.
For the pion, the same datasets are used as in previous JAM analyses~\cite{Barry:2018ort, Cao:2021aci, Barry:2021osv, Barry:2022aix, Barry:2023qqh, Good:2025nny}, namely, DY data from the CERN NA10~\cite{NA10:1985ibr} and Fermilab E615~\cite{Conway:1989fs} experiments (117 data points), and LN electroproduction data from H1~\cite{H1:2010hym} and ZEUS~\cite{ZEUS:2002gig} at HERA (108 points).
For consistency with the pion analyses, in which only data with $x_F=x_1-x_2 \geq 0$ were fitted (to minimize rescattering effects~\cite{NA10:1985ibr}), we limit the NA3 experimental coverage to the same kinematics, with the $x_F$ cuts based on the mid-points of the bins of $x_1$ and $x_2$.
In the range of $0.4 \leq x_1 \leq 1$, with a corresponding range of $0.1 \lesssim x_2 \lesssim 0.7$, the cross sections are dominated by the fusion of a valence $\bar u$ quark in the beam with a $u$ quark in the target nucleus.

To determine the strange quark PDF in the kaon requires additional constraints, for which we use lattice QCD data from ETMC~\cite{ExtendedTwistedMass:2024kjf} on several low moments of quark and gluon PDFs, defined by
\begin{eqnarray}
    \langle x^n \rangle_q  
    &=& \int_0^1 \dd{x} x^n \big[ q(x,\mu)  + (-1)^{n+1} \bar q(x,\mu) \big],
    \\
    \langle x^n \rangle_g 
    &=& \int_0^1 \dd{x} x^n g(x,\mu),\ \ \ n\ {\rm odd},
\end{eqnarray}
and computed at the physical pion mass with controlled continuum extrapolation.
These include the momentum fractions for the $q=u, d, s, c$ and $g$ distributions at a scale of $\mu=2$~GeV for both the pion and kaon.

For the $u$ and $s$ PDFs in the kaon and the $u$ PDF in the pion we also include lattice data~\cite{Alexandrou:2021mmi} on the $n=2$ and 3 moments, computed with connected contributions only at a pion mass of 260~MeV and at a single lattice spacing (see Table~\ref{t.LQCDmoments}).
To account for the potential missing systematic effects~\cite{ExtendedTwistedMass:2024kjf}, we inflate the uncertainties on the higher moments by a factor of two.
(Inflating the uncertainties by a factor of 3 has a minimal impact on our results.)
We also note the excellent agreement for the $\langle x \rangle_u$ moment in the pion between the connected$+$disconnected calculation in the continuum limit from Ref.~\cite{ExtendedTwistedMass:2024kjf} and the connected-only, single-lattice spacing result from Ref.~\cite{Oehm:2018jvm}.
This suggests that use of the higher moments from the single-lattice spacing calculations of Ref.~\cite{Alexandrou:2021mmi} should provide reliable constraints on the physical PDFs in our analysis.

\begin{table}[b]
\centering
\caption{Moments $\langle x^n \rangle$, with $n=1,2,3$, of kaon and pion PDFs at $\mu=2~{\rm GeV}$ from ETMC~\cite{ExtendedTwistedMass:2024kjf, Alexandrou:2021mmi} and the current JAM analysis.}
\begin{tabular}{c|ll|ll}
\hline
 & \multicolumn{2}{c|}{\it kaon} & \multicolumn{2}{c}{\it ~~pion} \\
Moment & ~\,ETMC & ~~\,JAM & ~~ETMC & ~~~JAM 
\\ \hline
$\langle x \rangle_u$ & 
~$0.269(9)$ & 
~$0.269(7)$ & 
~$0.249(28)$ & 
~$0.256(10)$ \\
$\langle x \rangle_d$ & 
~$0.059(9)$ & 
~$0.051(8)$ & 
~$0.249(28)$ & 
~$0.256(10)$ \\
$\langle x \rangle_s$ & 
~$0.339(11)$ & 
~$0.340(11)$ & 
~$0.036(15)$ & 
~$0.040(10)$ \\
$\langle x \rangle_c$ & 
~$0.028(21)$ & 
~$0.0071(3)$ & 
~$0.013(16)$ & 
~$0.0092(6)$ \\
$\langle x \rangle_g$ & 
~$0.422(67)$ & 
~$0.333(17)$ & 
~$0.402(53)$ &
~$0.439(28)$ \\ \hline
$\langle x^2 \rangle_u$ &
~$0.096(3)$ &
~$0.089(4)$ &
~$0.110(14)$ & 
~$0.090(4)$ \\
$\langle x^2 \rangle_s$ &
~$0.139(2)$ & 
~$0.137(4)$ & 
~~~~~--- & 
~~~~~--- \\ \hline
$\langle x^3 \rangle_u$ &
~$0.033(6)$ & 
~$0.050(3)$ & 
~$0.024(18)$ & 
~$0.052(2)$ \\
$\langle x^3 \rangle_s$ &
~$0.073(5)$ &
~$0.082(5)$ &
~~~~~--- & 
~~~~~--- \\ \hline
\end{tabular}
\label{t.LQCDmoments}
\end{table}

{\it Methodology---} 
For the simultaneous analysis of the pion and kaon PDFs, we perform a Bayesian inference using a Monte Carlo framework developed in recent JAM analyses of proton PDFs~\cite{Cocuzza:2021cbi, Cocuzza:2021rfn, Cocuzza:2022jye, Anderson:2024evk, Cocuzza:2025qvf}.
Following most previous global QCD analyses of pion PDFs~\cite{Owens:1984zj, Aurenche:1989sx, Sutton:1991ay, Gluck:1991ey, Gluck:1999xe, Wijesooriya:2005ir, Aicher:2010cb, Novikov:2020snp, Bourrely:2022mjf, Kotz:2023pbu, Kotz:2025lio}, we parametrize the PDF of flavor $f$ in the $K^-$ or $\pi^-$ at the input scale $\mu_0 = m_c = 1.28$~GeV using the form
\begin{equation}
f(x,\mu_0;\boldsymbol{a}) 
= N\, x^{\alpha}(1-x)^{\beta} (1 + \gamma \sqrt{x} + \delta x).
\label{eq.parametrization}
\end{equation}
For the pion, we assume charge symmetry, so that $\bar u^{\pi^-}\! = d^{\pi^-}\! = u^{\pi^+}\! = \bar d^{\pi^+}$.
In the kaon, charge symmetry implies that $\bar{u}^{K^-}\! = u^{K^+}$ and $s^{K^-}\! = \bar{s}^{K^+}$; in the valence sector, however, we allow $\bar{u}^{K^-}\! \neq s^{K^-}$.
Baryon number and momentum sum rules are imposed parametrically on the normalizations of the valence quarks and sea quarks, respectively, for both pions and kaons.

Because of the paucity of data on the kaon sea quark and gluon PDFs at small $x$, we equate the shapes for these distributions to those in the pion, $q_{\rm sea}^{K^-}\! \propto q_{\rm sea}^{\pi^-}$ and $g^{K^-}\! \propto g^{\pi^-}$, while allowing the normalizations to vary.
We also set the $\gamma$ and $\delta$ parameters of the pion sea quark and gluon PDFs to zero (implying the same for the kaon), as these do not improve the description of the data.
This gives 9 pion PDF parameters, 9 kaon parameters, and 6 data normalization parameters.
In addition, we fit the cutoff mass for the $p\to \pi^+n$ splitting function for the LN observables~\cite{McKenney:2015xis, Barry:2018ort}, giving a total of 25 free parameters to be determined in the fit.

In some early fits of pion and kaon PDFs~\cite{Gluck:1997ww}, the valence strange and $\bar u$ distributions in $K^-$ were related to the valence PDF in the pion via a constituent quark model inspired relation,
    $s_v^{K^-}(x) = 2 \bar{u}_v^{\pi^-}(x) - \bar{u}_v^{K^-}(x)$.
We explored using this constraint, but found it to be too 
restrictive to obtain good agreement between data and theory.

\begin{figure}[t]
    \hspace{-0.45cm}\includegraphics[width=1.04\linewidth]{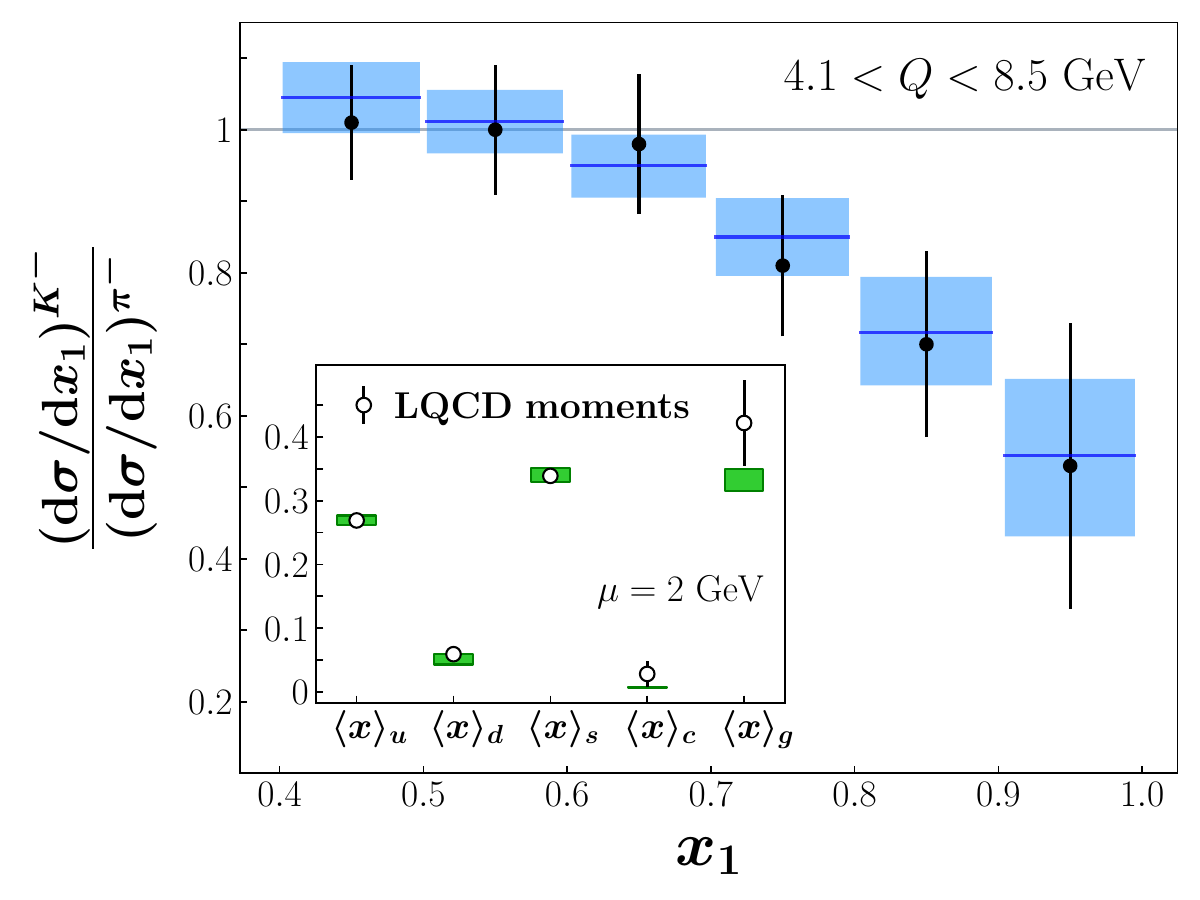}
    \vspace*{-0.3cm}
    \caption{Ratio of $K^-$ to $\pi^-$ DY cross sections from the NA3 experiment \cite{NA3:1980fhh} (black circles) compared with the JAM analysis (blue bars). The inset compares kaon PDF moments from JAM (green bars) with lattice QCD (LQCD) results from the ETMC (open circles) \cite{ExtendedTwistedMass:2024kjf}, shown for a 68\% CI.}
\label{f.datatheory}
\end{figure}

{\it QCD analysis---}
We collect $\mathcal{O}(1000)$ Monte Carlo samples of the Bayesian posterior distribution to obtain a representative set of PDF replicas for a reliable uncertainty quantification.
In Fig.~\ref{f.datatheory} the ratio of the $K^-$ to $\pi^-$ cross sections $\dd\sigma/\dd{x_1}$ from the NA3 experiment~\cite{NA3:1980fhh} is compared with our results.
The NA3 data are fitted very well, with a relatively small $\chi^2/N_{\rm dat}=0.08$, reflecting the large statistical uncertainties of the experiment.
For the pion-induced DY and LN datasets we find $\chi^2/N_{\rm dat} = 0.78$ and 0.87, respectively, for a total $\chi^2/N_{\rm dat} = 0.80$ for the experimental data.

Also shown in Fig.~\ref{f.datatheory} (see inset) is a comparison with the lattice QCD momentum fractions $\langle x \rangle$ carried by $u$, $d$, $s$ and $c$ quarks and gluons in the kaon at $\mu = 2$~GeV.
The agreement with the lattice results is very good, with $\chi^2/N_{\rm dat} = 0.88$ and 0.30 for the kaon (9 moments) and pion (6 moments), respectively, and $\chi^2/N_{\rm dat} = 0.65$ for the lattice data overall (15 data points).
For the combined experiment and lattice fit we have the total $\chi^2/N_{\rm dat} = 0.79$.
Note that while the momentum sum rule is imposed exactly in our QCD analysis, in the lattice simulation \cite{ExtendedTwistedMass:2024kjf} it is not used as an input, and subsequently for the kaon the sum rule is slightly overestimated, though still within $\approx 1\sigma$ uncertainty.
In the pion sector, however, the lattice calculated \cite{ExtendedTwistedMass:2024kjf} momentum sum rule is consistent with 1 within uncertainties, and accordingly our analysis finds good agreement with the momentum fraction of the gluon as well as with the experimental data.

Interestingly, while the momentum fractions carried by $u+\bar u$ quarks in the pion and kaon are comparable,
the momentum carried by strange quarks in the kaon is $\approx 30\%$ larger.
Similarly, the higher moments of the strange quark PDF in the kaon are larger than the corresponding $u$ moments, which pushes the $s^{K^-}$ PDF to higher values of $x$ than $\bar u^{K^-}$ or $\bar u^{\pi^-}$.
For the gluon, our analysis prefers a smaller fraction in the kaon than in the pion,
     $\langle x\rangle_g^{K} \lesssim \langle x\rangle_g^{\pi}$,
although the lattice results are comparable for kaons and pions, within much larger uncertainties than for quarks.
This result has consequences for the mass decomposition of the mesons \cite{Ji:1994av, Ji:1995sv, Metz:2020vxd}, where our results indicate that $\approx\!1/4$ (1/3) of the kaon (pion) mass is gluonic in origin at $\mu = 2$~GeV, reflecting the enhanced contribution from the heavier $s$ quark in the~kaon.

Note that since the $n=2$ and $n=3$ lattice moments were computed from connected contributions only~\cite{Alexandrou:2021mmi}, our results agree more closely with $\langle x^2 \rangle_q$, where the disconnected pieces cancel, than with $\langle x^3 \rangle_q$.
To estimate the error that this may introduce into the analysis, we repeated the fits with the lattice $\langle x^3 \rangle_u^K$, $\langle x^3 \rangle_s^K$, and $\langle x^3 \rangle_u^\pi$ moments increased by 7\%, 4.6\%, and 6\%, respectively, reflecting the differences between the phenomenological valence-only and the valence+sea contributions, but found no discernible effects on the valence PDFs.

\begin{figure}[t]
    \centering
    \includegraphics[width=1\linewidth]{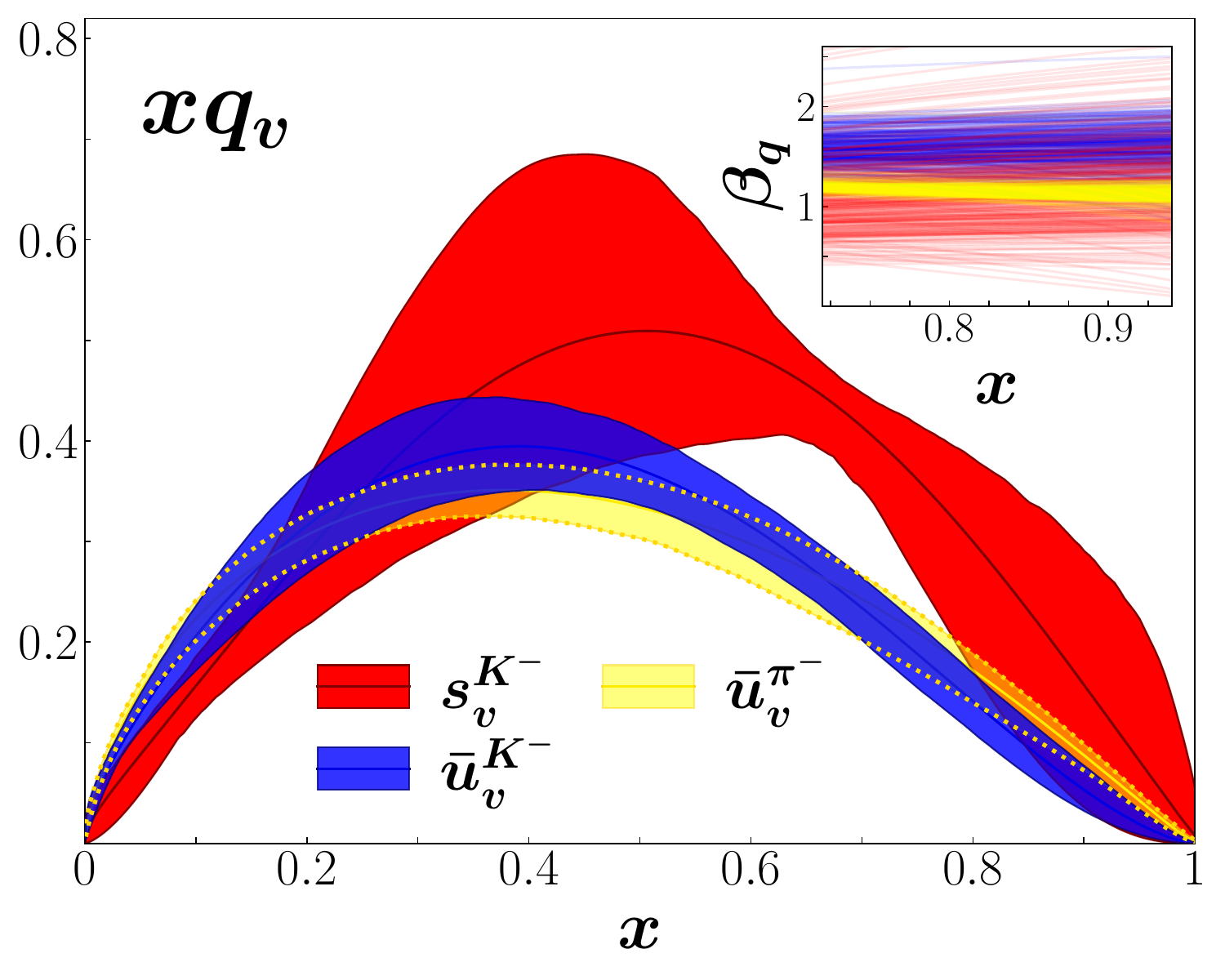}
    \vspace*{-0.6cm}
    \caption{Valence quark PDFs $x q_v$ in the $K^-$ for the $s$ (red) and $\bar u$ (blue) quarks, and the $\bar u$ in the $\pi^-$ (yellow), at the input scale $\mu = m_c$ for the 95\% CI. The inset shows the effective $\beta_q$ exponents at large $x$ as replicas.}
\label{f.pdfs}
\end{figure}

The resulting valence quark PDFs in the kaon and pion are shown in Fig.~\ref{f.pdfs} at the input scale $\mu=m_c$.
While the softening of the $\bar{u}_v^{K^-}$ PDF compared with $\bar{u}_v^{\pi^-}$ is expected from the shape of the NA3 DY data in Fig.~\ref{f.datatheory}, the behaviors of the $s$ and $\bar u$ valence quark PDFs in the kaon are strikingly different, with $s_v^{K^-} \gg \bar{u}_v^{K^-}$ at \mbox{$x \gtrsim 0.5$}.
A~harder $s_v^{K^-}$ distribution may be expected from considerations of the heavy quark limit~\cite{Neubert:1993mb}, where the heavy quark carries almost the entire momentum of the heavy meson, approaching a $\delta$ function at $x=1$ as $m_q \to \infty$ (see also discussions of $c$-quark distributions in $D$ mesons, where similar trends are expected~\cite{Paiva:1996dd, Melnitchouk:1997ig}).
While the hierarchy between the $s^{K^-}$, $\bar{u}^{\pi^-}$, and $\bar{u}^{K^-}$ PDFs has previously been observed in Ref.~\cite{Bourrely:2023yzi} and in some phenomenological model calculations~\cite{Bednar:2018mtf, Hutauruk:2016sug}, our analysis is the first model-independent, data-driven confirmation of the hierarchy strictly based on rigorously proved QCD factorization theorems, and unbiased by specific parametrizations chosen for the PDFs.

A comparison of the high-$x$ behavior of the valence quark PDFs is illustrated in the inset of Fig.~\ref{f.pdfs}, where we show replicas of the effective $\beta$ exponents for the valence kaon and pion PDFs, defined as
$\beta_q(x,\mu) = \partial\log\big|q_v(x,\mu)\big|/\partial \log(1-x)$. 
Clearly, the $x \to 1$ falloff for $\bar{u}_v^{K^-}$ is faster than for the
$\bar{u}_v^{\pi^-}$, with effective exponents 
    $\beta_{\bar{u}}^{K^-} = 1.6(2)$
and
    $\beta_{\bar{u}}^{\pi^-} = 1.16(4)$
for $x = 0.7-0.95$, the latter which is consistent with the earlier JAM analysis~\cite{Barry:2021osv}.
The strange valence quark has a wide spread of effective $\beta$ over the replicas, consistent with $\beta_s^{K^-} = 1.2(4)$, reflecting greater uncertainty in the strange quark's large-$x$ behavior. 
The uncertainties for those values are given by the 68\% credible interval (CI).

\begin{figure}[t]
    \centering
    \includegraphics[width=1\linewidth]{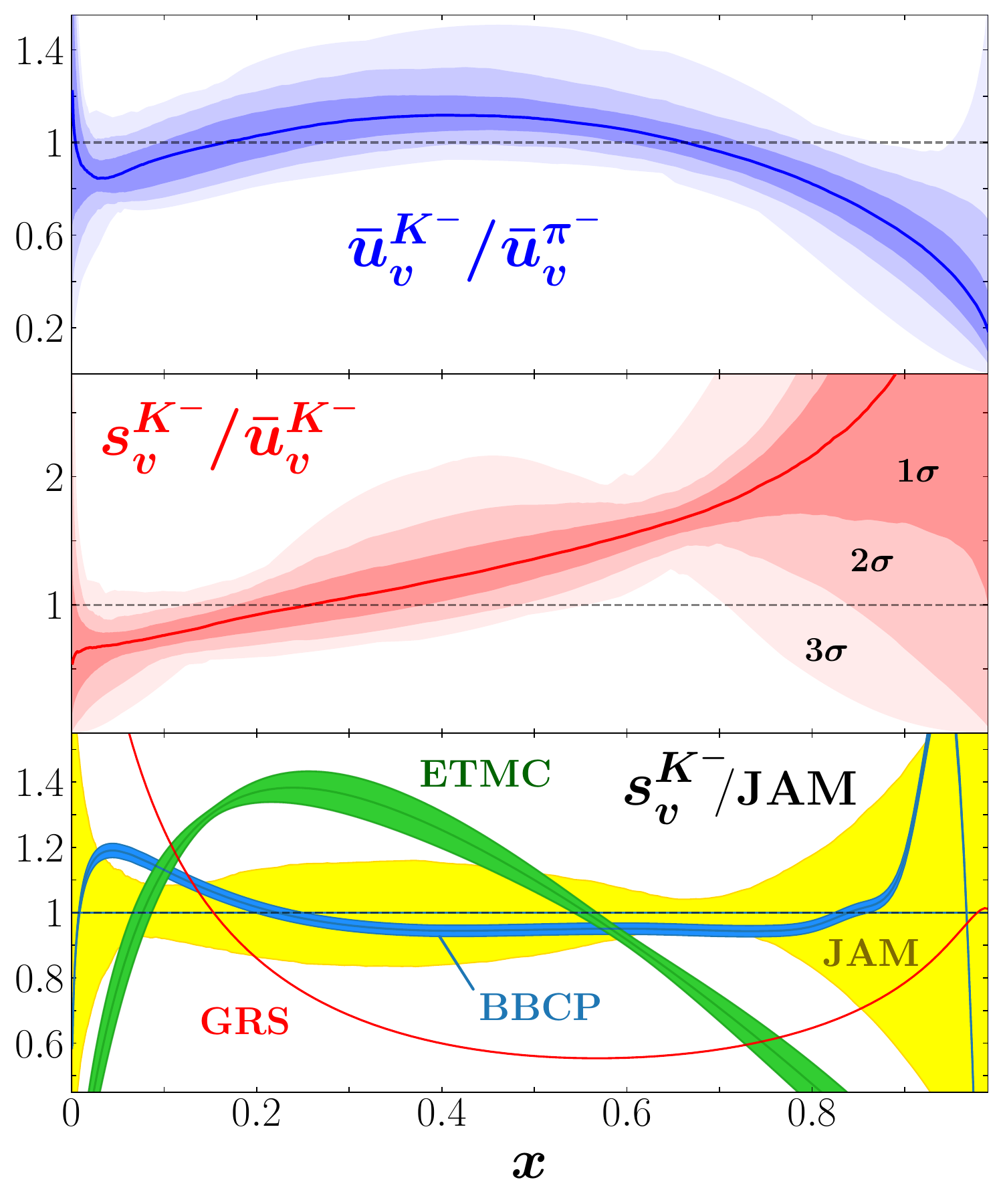}
    \vspace*{-0.6cm}
    \caption{Ratios of the $\bar{u}_v^{K^-}$ to $\bar{u}_v^{\pi^-}$ PDFs (upper panel) and $s_v^{K^-}$ to $\bar{u}_v^{K^-}$ PDFs (middle panel) for $1\sigma$ (darkest shading), $2\sigma$, and $3\sigma$ (lightest shading) CIs. Ratios of $s_v^{K^-}$ to the mean value of the JAM results (bottom panel) are shown for 68\% CIs for this analysis (yellow) compared with results from BBCP~\cite{Bourrely:2023yzi} (blue), GRS~\cite{Gluck:1997ww} (red), and  ETMC~\cite{Alexandrou:2021mmi} (at $\mu=2~{\rm GeV}$, green).}
    \label{f.ratios}
\end{figure}

The relative differences between the $x$ dependence of the valence quark PDFs can be more clearly illustrated via their ratios, as shown in Fig.~\ref{f.ratios}.
The $\bar{u}_v^{K^-}/\bar{u}_v^{\pi^-}$ ratio is consistent with unity for $x \lesssim 0.6$, above which it mirrors the downward trend of the NA3 data.
The ratio of the valence quark PDFs in the kaon, $s_v^{K^-}/\bar{u}_v^{K^-}$, has a stronger $x$ dependence, increasing monotonically with $x$.
At $x = 0.8$, the $s_v^{K^-}$ distribution is $\approx 2$ times as large as the $\bar{u}_v^{K^-}$ PDF within $1\sigma$, indicating the dominance of the strange quark PDF in the kaon at large $x$.

The $s_v^{K^-}$ PDF is also compared in Fig.~\ref{f.ratios} with experimental-only analyses from BBCP~\cite{Bourrely:2023yzi} and GRS~\cite{Gluck:1997ww} and the lattice moment-only analysis from ETMC~\cite{Alexandrou:2021mmi} as a ratio to the mean value of $s_v^{K^-}$ in the present JAM analysis.
The BBCP fit, which is based on a specific model-inspired PDF parametrization and includes $J/\psi$ data using nonrelativistic QCD with input on long-distance matrix elements~\cite{Hsieh:2021yzg}, is consistent with our result over most of the $x$ range, while significant differences are observed for the GRS and ETMC extractions.

\begin{figure}[t]
    \centering
    \includegraphics[width=1.0\linewidth]{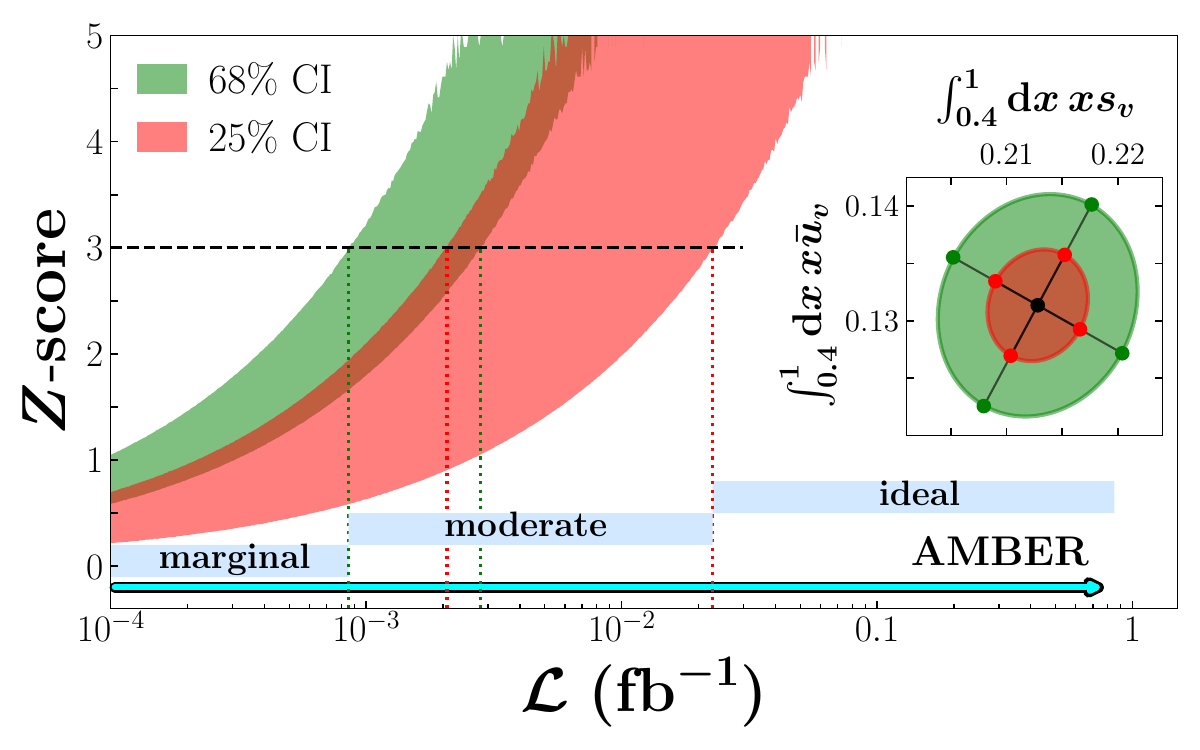}
    \vspace*{-0.6cm}
    \caption{$Z$-score from the $t$-comparison statistic as a function of luminosity for the proposed kaon-induced AMBER experiment \cite{Denisov:2018unj}. Luminosities above $2 \times 10^{-2}$~fb$^{-1}$ would be ideal for $3\sigma$ discrimination. The inset shows the distributions of the truncated $s_v^{K^-}$ and $\bar{u}_v^{K^-}$ moments.}
    \label{f.impact}
\end{figure}

{\it Outlook---}
The behavior of the valence and sea quark PDFs in the kaon will be further elucidated by the AMBER experiment~\cite{Denisov:2018unj} at CERN, which plans to measure the first kaon-induced DY cross sections in nearly 50 years.
To assess the impact of these measurements, we estimate in Fig.~\ref{f.impact} the discriminating power that kaon DY cross sections at AMBER kinematics would have on the current uncertainties on the kaon valence PDFs.
Specifically, we perform discrimination tests of the PDF samples that sit at the boundaries of 25\% and 68\% CIs against the mean values using log-likelihood ratio tests mapped into $Z$-scores as a function of machine luminosity.
The likelihood ratios are computed for the truncated values of the $x$-weighted moments of $\bar{u}_v^{K^-}$ and $s_v^{K^-}$.
While marginal impact is found for luminosities below $10^{-3}~{\rm fb}^{-1}$, and moderate impact is observed in the range of $10^{-3}$ to $10^{-2}$~fb$^{-1}$, an ideal discrimination beyond $3\sigma$ can be achieved for luminosities above $\approx 2 \times 10^{-2}$~fb$^{-1}$.
Availability of $K^+$ beams, in addition to $K^-$, would significantly enhance the ability to resolve the different PDF flavors in kaons.

Other experiments, 
at Jefferson Lab with 11~GeV or upgraded 22~GeV electron beams, or at the 
Electron-Ion Collider (EIC), could access the kaon PDFs via the Sullivan process~\cite{Sullivan:1971kd} through detection of a $\Lambda$ hyperon in coincidence with the scattered electron, $e p \to e \Lambda X$.
The Jefferson Lab experiments would kinematically access the valence region, having substantial overlap with AMBER, while the EIC would be sensitive to sea quark and gluon distributions at small $x$.
Constraints on the kaon chiral splitting function, which will be needed to interpret these types of experiments, can be estimated from previous $pp \to \Lambda X$ measurements 
\cite{Blobel:1978yj}.

Additional constraints on strangeness in the kaon could come from kaon-induced charged-current charm meson production, which can access the strange quark through a charged $W$-boson exchange, $e p \to \nu D X$, as well as tagged semi-inclusive DIS, $e p \to e \Lambda K X$, which will require knowledge of the kaon fragmentation functions from global analyses \cite{Moffat:2021dji, Anderson:2024evk, deFlorian:2017lwf, AbdulKhalek:2022laj}.
While challenging, such next-generation experiments could be instrumental in providing for the first time a full picture of the full quark flavor decomposition of the simplest strange hadron.

The replicas from our QCD analysis, which we refer to as {\tt JAM25kaon} and {\tt JAM25pion}, are available upon request.

\vspace*{0.3cm}
{\it Acknowledgments---} 
We thank J.~McKenney and \mbox{J.-C.~Peng} for helpful discussions and communications. This work was supported by the U.S.~Department of Energy, Office of Science, Office of Nuclear Physics, contract no.~DE-AC02-06CH11357, and the Scientific Discovery through Advanced Computing (SciDAC) award {\it Femtoscale Imaging of Nuclei using Exascale Platforms} (P.B.), by the U.S. Department of Energy contract No. DE-FG02-03ER41260 (C.J.), and by the U.S. Department of Energy contract No.~DE-AC05-06OR23177, under which Jefferson Science Associates, LLC operates Jefferson Lab (W.M. and N.S.). The work of N.S. was supported by the DOE, Office of Science, Office of Nuclear Physics in the Early Career Program. 
F.S was funded in part by the Deutsche Forschungsgemeinschaft (DFG, German Research Foundation) as part of the
CRC 1639 NuMeriQS – project No. 511713970.
We acknowledge the Quark-Gluon Tomography (QGT)
Topical Collaboration funded by the U.S. Department of Energy, Office of Science,
Office of Nuclear Physics with Award DE-SC0023646.

\bibliography{ref.bib}

\end{document}